\documentclass{aa}
\usepackage{graphicx}
\usepackage{epsfig}
\usepackage{natbib}
\usepackage{txfonts}
\bibpunct{(}{)}{;}{a}{}{,}
\begin{document}

\title{The ELODIE survey for northern extra-solar planets\\
II: A Jovian planet on a long-period orbit around GJ\,777\,A\thanks{
Based on observations made with the {\footnotesize ELODIE} echelle spectrograph mounted on the 1.93--m Telescope 
at the Observatoire de Haute-Provence ({\footnotesize CNRS}) and with the {\footnotesize AFOE} spectrograph mounted on the 
1.5--m Telescope at the Fred Lawrence Whipple Observatory (SAO).} \thanks{The 
{\footnotesize ELODIE} and {\footnotesize AFOE} measurements discussed in this paper are only available in 
electronic form at the {\footnotesize CDS} via anonymous ftp to {\tt 
cdsarc.u-strasbg.fr (130.79.125.5)} or via 
{\tt http://cdsweb.u-strasbg.fr/Abstract.html}}}

\author{D.~Naef\inst{1} \and M.~Mayor\inst{1} \and S.G.~Korzennik\inst{2} \and D.~Queloz\inst{1} \and S.~Udry\inst{1} \and
P.~Nisenson\inst{2} \and R.W.~Noyes\inst{2} \and T.M.~Brown\inst{3} \and J.L.~Beuzit\inst{4} \and C.~Perrier\inst{4} \and J.P.~Sivan\inst{5}
}

\institute{Observatoire de Gen\`eve, 51 ch. des Maillettes, 
CH--1290 Sauverny, Switzerland \and
Harvard-Smithsonian Center for Astrophysics, 60 Garden Street, Cambridge, MA-02138, USA \and
High Altitude Observatory/National Center for Atmospheric Research, 3450 Mitchell Lane, Boulder, CO-80307, USA \and
Laboratoire d'Astrophysique, Observatoire de Grenoble, 
Universit\'e J. Fourier, BP 53, F--38041 Grenoble, France \and
Observatoire de Haute-Provence, F--04870 St-Michel L'Observatoire, France
}

\offprints{Dominique Naef,
\email{Dominique.Naef@obs.unige.ch}}

\date{Received / Accepted}

\abstract{We present radial-velocity measurements obtained with the {\footnotesize ELODIE} and {\footnotesize AFOE} spectrographs
for \object{{\footnotesize GJ}\,777\,A} (\object{{\footnotesize HD}\,190360}), a metal-rich 
($[$Fe/H$]$\,=\,0.25) nearby ($d$\,=\,15.9\,pc) star in a stellar binary system.
A long-period low radial-velocity amplitude variation is detected revealing the presence of a Jovian planetary companion. Some of the 
orbital elements remain weakly constrained because of the smallness of the signal compared to our instrumental precision. 
The detailed 
orbital shape is therefore not well established. We present our best fitted orbital solution: an eccentric 
($e$\,=\,0.48) 10.7--year orbit. The minimum mass of the companion is 1.33\,M$_{\rm Jup}$.
}

\titlerunning{The ELODIE survey for northern extra-solar planets II}
\authorrunning{D.~Naef et al.}
\maketitle

\section{Introduction}\label{intro}
The {\em {\footnotesize ELODIE} survey for northern extra-solar planets} is a programme aiming at detecting planetary companions 
around Solar-type stars with the {\footnotesize ELODIE} echelle spectrograph \citep{Baranne96} mounted on the 193--cm Telescope 
of the Observatoire de Haute-Provence. 
Details about the programme and the surveyed sample have already been presented \citep{Mayorcssss, PerrierELODIE1}. 
Several planets have been detected by this survey \citep[see][ and references therein]{PerrierELODIE1}.

Four extra-solar planets have been discovered or codiscovered by the the Advanced Fiber-Optic Echelle spectrometer
\citep[{\footnotesize AFOE, }][]{Brown1994} planet search team \citep{KorzennikCSSSS}: \object{$\rho$\,CrB}\,b \citep{Noyesrhocrb}; 
\object{Ups\,And}\,c and d \citep{Butlerupsand} and \object{{\footnotesize HD}\,89744}\,b \citep{Korzennik}.

In 1997, our two teams decided to start a collaboration on a sample of about 20 stars present in both observing 
lists and having a similar radial-velocity precision of the order of 10\,m\,s$^{\rm -1}$. The detection of the 
low-amplitude radial-velocity variability of \object{{\footnotesize GJ}\,777\,A} and the characterisation of the planetary companion 
responsible for these variations are the first results of our common effort.

A preliminary {\footnotesize ELODIE} orbital solution was presented in \citet{Udrywash}.
In Sect.~\ref{RV}, we present our combined radial-velocity data and the updated orbital solution. 
We also examine and rule out alternative explanations for the origin of 
the observed radial-velocity variation.

\begin{table}[th!]
  \caption{
  \label{startab}
  Observed and inferred stellar parameters for GJ\,777\,A.
}
\begin{tabular}{llr@{  $\,\pm\,$  }l}
\hline
HD                          &                     & \multicolumn{2}{c}{190360}\\
HIP             	    &                     & \multicolumn{2}{c}{98767}\\
$Sp.\,Type$                 &                     & \multicolumn{2}{c}{G6IV}\\
$m_{\rm V}$                 &                     & \multicolumn{2}{c}{5.73}\\
$B-V$                       &                     & 0.749     & 0.001\\
$\pi$                       & (mas)               & 69.92     & 0.62\\
$Distance$                  & (pc)                & 15.89     & 0.16\\
$\mu _{\alpha}\cos(\delta)$ & (mas yr$^{\rm -1}$) & 683.32    & 0.42\\
$\mu _{\delta}$             & (mas yr$^{\rm -1}$) & $-$524.06 & 0.51\\
$M_{\rm V}$                 &                     & \multicolumn{2}{c}{4.72}\\
$B.C.$                      &                     & \multicolumn{2}{c}{$-$0.1175}\\
$L$                         & (L$_{\sun}$)        & \multicolumn{2}{c}{1.13}\\
$T_{\rm eff}$               & ($\degr$K)          & 5590      & 50\\
$[$Fe/H$]$                  &                     & 0.25      & 0.05\\
$\log g$                    & (cgs)               & 4.48      & 0.15\\
$\xi _{\rm t}$              & (km\,s$^{-1}$)	  & 1.06      & 0.10\\
$\log R^{'}_{\rm HK}$       &                     & \multicolumn{2}{c}{$-$5.05}\\
$P_{\rm rot}$               & (d)                 & \multicolumn{2}{c}{38}\\
$age({\rm HK})$	 	    & (Gyr)               & \multicolumn{2}{c}{6.7}\\
$v\sin i$                   & (km\,s$^{-1}$)      & \multicolumn{2}{c}{$<$1}\\
$R_{\ast}$                  & (R$_{\sun}$)        & \multicolumn{2}{c}{1.13}\\
$M_{\ast}$                  & (M$_{\sun}$)        & \multicolumn{2}{c}{0.96}\\[0.05cm]
\hline
\end{tabular}
\end{table}

\section{Stellar properties}\label{star}

The main stellar properties of \object{{\footnotesize GJ}\,777\,A} are listed in Table~\ref{startab}.
Spectral Type, $m_{\rm V}$, $B-V$, $\pi$, $\mu _{\alpha}\cos(\delta)$ and $\mu _{\delta}$ are from the 
{\footnotesize HIPPARCOS} Catalogue \citep{ESA97}. The effective temperature $T_{\rm eff}$, the surface gravity $\log g$, 
the metallicity $[$Fe/H$]$ and  the microturbulence velocity $\xi_{\rm t}$ are taken from \citet{Santosstat}. 
$M_{\ast}$ is the stellar mass derived from these atmospheric 
parameters using the Geneva evolutionary models \citep{Schaerer93}. The bolometric correction $B.C.$ is 
computed from the effective temperature with the calibration in \citet{Flower96}. 

The star does not appear to be active. We do not see any trace of emission in the core of the $\lambda$\,3968.5\,\AA\hspace{1mm}
\ion{Ca}{ii} H line on our {\footnotesize ELODIE} coadded spectra. A $\log R^{'}_{\rm HK}$\,=\,$-$5.05
chromospheric activity indicator is computed from the $S$ chromospheric flux indexes in \citet{Duncan91}. 
The rotation period $P_{\rm rot}$ and $age({\rm HK})$ are derived from the activity indicator using the calibrations in 
\citet{Noyes84} and \citet{Donahue93}\footnote{also quoted in \citet{Henry96}}, respectively.  
The rotational broadening measured for \object{{\footnotesize GJ}\,777\,A} from {\footnotesize ELODIE} cross-correlation 
function widths is very small. It only allows us to derive an upper limit for the projected rotational velocity: 
$v\sin i$\,$<$\,1\,km\,s$^{\rm -1}$. The radius $R_{\ast}$ is computed from $L$ and $T_{\rm eff}$. 
With a measured {\footnotesize HIPPARCOS} photometric scatter of 7\,mmag, \object{{\footnotesize GJ}\,777\,A} is flagged as a 
constant star.

\object{{\footnotesize GJ}\,777\,B}, a 14.4--mag M4.5 companion about 3\arcmin\,away, is mentioned in the 
{\em Catalogue of Nearby Stars} \citep{Gliese}. Recent proper motions determinations by \citet{Bakos} give 
$\mu _{\alpha}\cos(\delta)$\,=\,737\,mas\,yr$^{\rm -1}$ and $\mu _{\delta}$\,=\,$-$551\,mas\,yr$^{\rm -1}$ for this object. 
These values are close to the proper motions quoted in Table~\ref{startab} for \object{{\footnotesize GJ}\,777\,A}, thus these two 
stars probably form a physical pair. The separation between A and B computed from the revised coordinates in \citet{Bakos} is 
$\rho$\,=\,177\arcsec. The minimum semi-major axis, $a_{\rm min}=0.5\rho/\pi$, is 1266\,AU for the \object{{\footnotesize GJ}\,777\,A}B 
binary. This leads to a minimum period of the order of 40\,000\,yr (assuming $M_{\rm A}$\,=\,0.96\,M$_{\sun}$ 
and $M_{\rm B}$\,=\,0.3\,M$_{\sun}$).

\begin{figure}
  \psfig{file=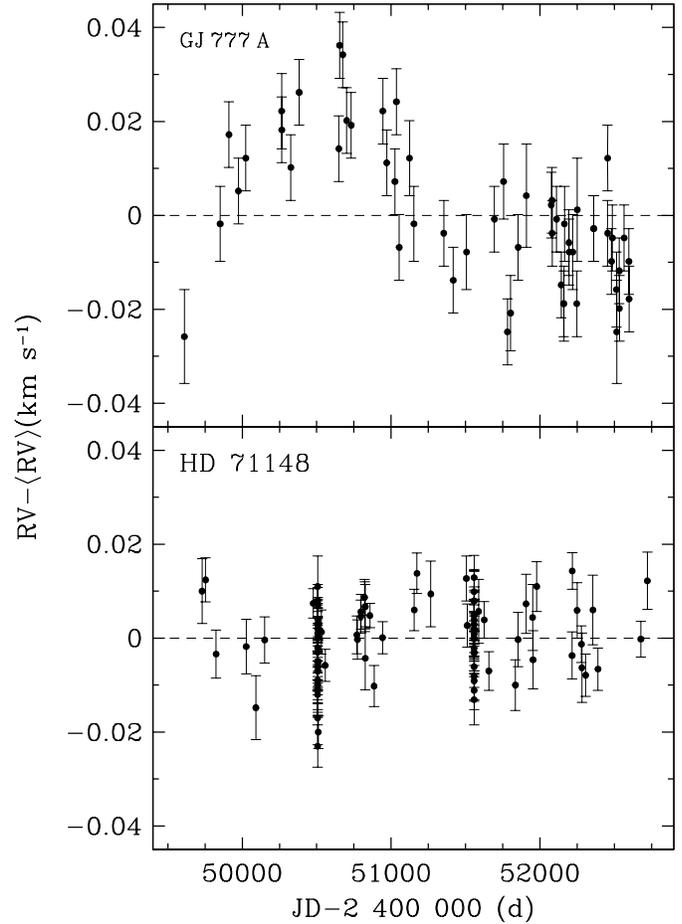,width=\hsize}\\
  \caption{
  \label{residuals} ELODIE data. {\bf TOP}: RV$-\langle RV\rangle$ for GJ\,777\,A. {\bf Bottom}: Residuals of HD\,71148,    
   a radial-velocity constant G5V star. The residuals slope is compatible with zero whereas a clear trend is seen for GJ\,777\,A. 
  This demonstrates that the GJ\,777\,A radial-velocity signal does not result from some undesirable instrumental effect.
  }
\end{figure}

\section{Radial-velocity data and orbital solution}\label{RV}

Between {\footnotesize JD}\,=\,2\,449\,611 (September 1994) and {\footnotesize JD}\,=\,2\,452\,598 (November 2002), 56 
{\footnotesize ELODIE} and 13 {\footnotesize AFOE} radial-velocity measurements have been gathered. The method used for 
extracting precise velocities from the observed spectra is described in \citet{Baranne96} and 
\citet{PerrierELODIE1} for {\footnotesize ELODIE} and in \citet{Noyesrhocrb} for {\footnotesize AFOE}. The mean 
radial-velocity uncertainty achieved for \object{{\footnotesize GJ}\,777\,A} is $\langle\epsilon_{\rm RV}\rangle$\,=\,7.5\,m\,s$^{\rm -1}$ 
with {\footnotesize ELODIE} and  $\langle\epsilon_{\rm RV}\rangle$\,=\,9.7\,m\,s$^{\rm -1}$ with {\footnotesize AFOE}. 
The $rms$ of the radial velocities is 15.2\,m\,s$^{\rm -1}$ (18.2\,m\,s$^{\rm -1}$) for {\footnotesize ELODIE} 
({\footnotesize AFOE}). The $\chi^{\rm 2}$ probability for hypothesis of constant velocities is less 
than 5$\cdot$10$^{\rm -4}$ for both instruments so 
\object{{\footnotesize GJ}\,777\,A} is clearly a radial-velocity variable.

\begin{table}[t!]
\caption{
\label{orbsol}
Combined ELODIE (E)\,+\,AFOE (A) orbital solution for Gl\,777\,A.
$\sigma_{\rm O-C}$ is the weighted rms of the residuals. 
$\chi^{\rm 2}_{\rm red}$ is the reduced $\chi^{\rm 2}$ value ($\chi^{\rm 2}/\nu$ where $\nu$ is the number of degrees of freedom).
}
\begin{tabular}{ll|r@{\,$\pm$\,}l}
\hline
\noalign{\vspace{0.05cm}}
$P$                                 & (days)                        & 3902      & 1758\\
$T$                                 & (JD$^{\dagger}$)              & 50\,557   & 89\\
$e$                                 &                               & 0.48      & 0.20\\
$\gamma$                            & (km\,s$^{\rm -1}$)            & $-$45.350 & 0.004\\
$w$                                 & ($\degr$)                     & 361       & 13\\
$K_{\rm 1}$                         & (km\,s$^{\rm -1}$)            & 0.020     & 0.003\\
$\Delta RV_{E-A}$                   & (km\,s$^{\rm -1}$)            & $-$45.143 & 0.003\\
$a_{\rm 1} \sin i$                  & (10$^{\rm -3}$AU)             & 6.3       & 2.7\\
$f_{\rm 1}(m)$                      & (10$^{\rm -9}$M$_{\sun}$)     & 2.22      & 0.94\\
\noalign{\vspace{0.025cm}}
\hline
\noalign{\vspace{0.025cm}}
$m_{\rm 2} \sin i$                  & (M$_{\rm Jup}$)               & 1.33       & 0.19\\
$a$                                 & (AU)                          & \multicolumn{2}{c}{4.8}\\
$d_{\rm min}$$^{\ddagger}$          & (AU)                          & \multicolumn{2}{c}{2.5}\\
$d_{\rm max}$$^{\natural}$          & (AU)       	            & \multicolumn{2}{c}{7.1}\\
\hline
$N$                                 &                               & \multicolumn{2}{c}{69}\\ 
$N_{E}$                             &                               & \multicolumn{2}{c}{56}\\
$N_{A}$                             &                               & \multicolumn{2}{c}{13}\\
$\sigma_{\rm O-C}$                  & (m\,s$^{\rm -1}$)             & \multicolumn{2}{c}{9.3}\\
$\sigma_{\rm O-C, E}$               & (m\,s$^{\rm -1}$)             & \multicolumn{2}{c}{9.1}\\
$\sigma_{\rm O-C, A}$               & (m\,s$^{\rm -1}$)             & \multicolumn{2}{c}{10.4}\\
$\chi^{\rm 2}_{\rm red}$            &                               & \multicolumn{2}{c}{1.69}\\[0.05cm]
\hline
\end{tabular}

$^{\dagger}$=\,{\footnotesize JD}$-$2\,400\,000 $^{\ddagger}$at periastron $^{\natural}$at apastron
\end{table}

\object{{\footnotesize GJ}\,777\,A} is a slowly rotating chromospherically quiescent star. The expected stellar-activity induced radial-velocity 
signal is therefore very low. Even so, we checked the stability of the line bisectors using the method presented in 
\citet{Queloz166435}. Our {\footnotesize ELODIE} bisectors were found to be constant without any trace of correlation 
with the radial velocities. The effect of long-period magnetic activity cycles on radial-velocity measurements is still 
not well known. The possibility that this kind of phenomenon is responsible for the observed long-period signal 
cannot be rejected for the moment. Accurate studies of this problem will be soon performed with the new 
{\footnotesize HARPS} spectrograph on the 3.6--m Telescope at {\footnotesize ESO}--La Silla Observatory, especially designed for 
high-precision radial-velocity measurements \citep{Pepemessenger}.

We also checked the radial-velocity signal against instrumental effects. Figure~\ref{residuals} shows 
a comparison between the observed \object{{\footnotesize GJ}\,777\,A} signal and the residuals for \object{{\footnotesize HD}\,71148}, 
a G5V non-variable star. The $rms$ of the 98 {\footnotesize HD}\,71148 residuals is 7.8\,m\,s$^{\rm -1}$ and their fitted slope is 
compatible with zero. The detected signal for \object{{\footnotesize GJ}\,777\,A} is 
not seen for this comparison star. 

\begin{figure}
  \psfig{file=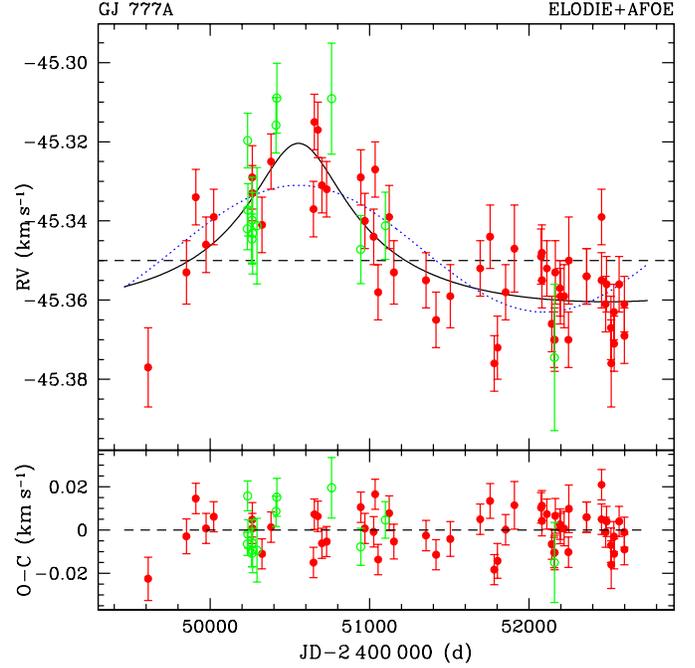,width=\hsize}\\
  \caption{
  \label{rvplot} {\bf TOP}: ELODIE (filled circles) and AFOE (open circles) temporal velocities. The fitted eccentric orbit is 
  shown (solid line). The radial-velocity maximum is not well covered. This probably tends to artificially increase the fitted 
  eccentricity. The fitted circular orbit (dotted line) is also plotted for comparison.
  {\bf Bottom}: Residuals to the fitted eccentric orbit.
  }
\end{figure}

The fitted orbital elements to the combined {\footnotesize ELODIE} and {\footnotesize AFOE} radial velocities are presented in 
Table~\ref{orbsol}. The {\footnotesize AFOE} velocity zero-point is arbitrary so we also adjust the radial-velocity offset 
between the two instruments: $\Delta RV$\,=\,$RV_{\rm ELODIE}-RV_{\rm AFOE}$. Some of the parameters are weakly 
constrained. This is specially the case for the period and the eccentricity. The time span of the 
{\footnotesize ELODIE} measurements is 2987\,d, a value smaller than the derived period. This causes a significant uncertainty 
in $P$. We can see in Fig.~\ref{rvplot} that the radial-velocity maximum is not perfectly covered. This probably 
increases the derived eccentricity. With these orbital elements and assuming 0.96\,M$_{\sun}$ for the primary star mass, we compute 
a minimum mass of 1.32\,M$_{\rm Jup}$ for \object{{\footnotesize GJ}\,777\,Ab} and an orbital semi-major axis of 4.8\,AU. 
As \object{{\footnotesize GJ}\,777\,A} is close to the Sun ($d$\,=\,15.9\,pc), its projected orbital semi-major axis is relatively large:
$a_{\rm 1}\sin i$\,$\simeq$\,0.4\,mas on the sky. This system is thus an easy target for the future interferometric astrometric 
facilities ({\footnotesize VLTI, SIM}).

We also computed the orbital solution using the {\footnotesize ORBIT} software described in \citet{Forveilleorbit}. The resulting 
orbital elements were in full agreement with the values in Table~\ref{orbsol}. This software also allowed us to check the fit 
errors by performing Monte--Carlo simulations. The output of these simulations are 68\% confidence level (i.e. 1\,$\sigma$) 
intervals for each parameter. As expected, the error in $P$ is largely asymmetric: $+$4700/$-$965\,d. This is also the case 
for the error in $e$: $+$0.23/$-$0.14. The Monte--Carlo errors in the other parameters are rather symmetric but 
sometimes slightly larger than the errors obtained from the fit. This is particularly the case for $T$ where the Monte--Carlo 
error is $\pm$\,118\,d.

We see from the Monte--Carlo simulations that a zero eccentricity is more than 3\,$\sigma$ away from the fitted value.
We nevertheless tried to fit a circular orbital solution to the data. The residuals value obtained for this latter solution is 
10.2\,m\,s$^{\rm -1}$ a value that is not much larger than the eccentric orbit value. From these facts, we conclude that a 
circular orbit cannot be excluded with the available data but it is rather unlikely.

\section{Concluding remarks}\label{conc}

The number of detected extra-solar planets resembling Jupiter slowly increases with 
the duration of the different radial-velocity planet searches. What would be in our sense a real Jupiter analog? 
It would be a planet with a period not too different from $P_{\rm Jup}$ (say for example: 
$P_{\rm Jup}$\,$\pm$\,25\%), a moderate eccentricity ($e<$0.2), a mass close to one Jupiter mass ($\pm$\,50\% for 
example). Moreover, this planet would have to be innermost gas giant in the system.
Table~\ref{analogs} lists the characteristics of the 
4 best Jupiter-analog candidates (planets with $P$\,$>$2500\,d and $m_{\rm 2}\sin i$\,$<$\,5\,M$_{\rm Jup}$): 
\object{$\epsilon$\,Eri}\,b \citep{Hatzesepseri}; \object{47\,UMa}\,c \citep{Fischer47umac}; 
\object{{\footnotesize GJ}\,777\,Ab} (this paper) and \object{55\,Cnc}\,d \citep{Marcy55cnc}. 
None of them matches all the criteria mentioned above.

The distributions of the orbital elements as well as the planet-mass distribution 
seem to be continuous so we do not see any reason why real Jupiter analogs would not eventually be detected.

\begin{table}
\caption{\label{analogs} Comparison between Jupiter and the extra-solar planets having
$P$\,$>$2500\,d and $m_{\rm 2}\sin i$\,$<$\,5\,M$_{\rm Jup}$. IGG is the 
Innermost Gas Giant flag (yes or no).
}
\begin{tabular}{lccccc}
\hline
\noalign{\vspace{0.05cm}}
Name                        & $P$       & $e$     & $m$ or $m\sin i$ & $a$      & IGG\\
                            & (yr)      &         & (M$_{\rm Jup}$)  & (AU)     &\\
\hline
\object{$\epsilon$\,Eri\,b} & 7         & 0.43    & 0.92             & 3.40     & y\\
\object{47\,UMa\,c}         & 7.1       & 0.1     & 0.76             & 3.73     & n\\
\object{GJ\,777\,Ab}        & 10.7      & 0.48    & 1.33             & 4.8      & y\\
\object{Jupiter}            & 11.86     & 0.05    & 1                & 5.2      & y\\
\object{55\,Cnc\,d}         & 14.7      & 0.16    & 4.05             & 5.9      & n\\
\noalign{\vspace{0.05cm}}
\hline
\end{tabular}
\end{table}

\object{{\footnotesize GJ}\,777\,Ab} is a planet in a stellar binary system. The number of extra-solar planets 
detected in multiple systems now approaches 20 (Eggenberger et al., in prep.). Although the sample of
such planets is not large, some properties are now emerging. Significant differences between planets 
in multiple systems and planets orbiting single stars are found \citep{ZuckerMazeh, Udrystat}. 
The study of planets in multiple systems is potentially very rich in information on formation scenarios 
and migration processes (Eggenberger et al., in prep.).

\begin{acknowledgements}
We acknowledge support from the Swiss National Research Found 
({\footnotesize FNRS}), the Geneva University and the French 
{\footnotesize CNRS}. We are grateful to the Observatoire de 
Haute-Provence for the generous time allocation. 
The {\footnotesize AFOE} research was funded in part by {\footnotesize NASA}
grants {\footnotesize NAG}\,5--7505 and {\footnotesize NAG}\,5--10854.  We are grateful to the staff of the Fred
Lawrence Whipple Observatory for help with the {\footnotesize AFOE} observations.
This research has made use 
of the {\footnotesize SIMBAD} database, operated at 
{\footnotesize CDS}, Strasbourg, France.
\end{acknowledgements}

\bibliographystyle{aa}
\bibliography{naef_elodieII}
\end{document}